\documentclass[3p,12pt]{elsarticle}
\usepackage[colorlinks=true,citecolor=blue,linkcolor=blue]{hyperref}
\usepackage{amsmath}
\usepackage{amssymb}
\usepackage{color}
\usepackage{graphicx}
\usepackage{longtable}
\usepackage{listings}
\usepackage{chngcntr}
\usepackage{wasysym}
\counterwithin{figure}{section}
\usepackage[dvipsnames]{xcolor}
\usepackage{array, booktabs, makecell}
\newcounter{bla}

\DeclareFixedFont{\ttb}{T1}{txtt}{bx}{n}{10} 
\DeclareFixedFont{\ttm}{T1}{txtt}{m}{n}{10}  
\DeclareFixedFont{\tti}{T1}{txtt}{it}{n}{10}  

\usepackage{color}
\definecolor{deepblue}{rgb}{0,0,0.5}
\definecolor{deepred}{rgb}{0.6,0,0}
\definecolor{deepgreen}{rgb}{0,0.5,0}
\usepackage{url}

\usepackage{listings}

\usepackage{titlesec}
\titleformat{\paragraph}[runin]
{\bfseries\scshape}{\theparagraph}{1em}{}

\newcommand{\lminus}{{B-L}}

\usepackage{lipsum}
\usepackage{blindtext}

\newcommand\pythonstyle{\lstset{
language=Python,
basicstyle=\ttm,
otherkeywords={self},             
keywordstyle=\ttb\color{deepblue},
emph={MyClass,__init__, uls},          
emphstyle=\ttb\color{deepred},    
stringstyle=\color{deepgreen},
frame=tb,                         
showstringspaces=false,            %
commentstyle=\tti,
morecomment=[s]{"""}{"""},
}}
\newcommand\bashstyle{\lstset{
language=bash,
basicstyle=\ttm,
otherkeywords={self},             
keywordstyle=\ttb\color{deepblue},
emph={uls-calc,uls-scan,uls-nest},          
emphstyle=\ttb\color{deepred},    
stringstyle=\color{deepgreen},
frame=tb,                         
showstringspaces=true,            %
commentstyle=\tti,
}}

\lstnewenvironment{python}[1][]
{
\pythonstyle
\lstset{#1}
}
{}

\lstnewenvironment{bash}[1][]
{
\bashstyle
\lstset{#1}
}
{}

\newcommand\pythoninline[1]{{\pythonstyle\lstinline!#1!}}

\usepackage{epiolmec}
\usepackage[dvipsnames]{xcolor}

\usepackage{lineno}

\usepackage{listings}
\usepackage{eso-pic}
\usepackage{epsfig,graphicx}
\usepackage{bm} 
\usepackage{slashed}
\newcommand*\cpp{C\kern-0.2ex\raisebox{0.4ex}{\scalebox{0.8}{+\kern-0.4ex+}}}

\usepackage{color}
\usepackage{pstricks}

\usepackage{fancyhdr}
\usepackage{textcomp}

\usepackage{xcolor} 
\usepackage{colortbl} 
 
\newcommand{\unit}{\leavevmode\hbox{\small1\kern-3.6pt\normalsize1}}
\def \ie{{\it i.e.}}
\def \eg{{\it e.g.}}

\def \GeV{{\mathrm{GeV}}}

\newcommand{\Mbh}{M_{\rm BH}}
\newcommand{\gs}{g_\ast}
\newcommand{\gss}{g_{\ast,\, s}}

\newcommand{\equaref}[1]{Eq.~(\ref{#1})}

\newcommand{\figref}[1]{Fig.~\ref{#1}}

\newcommand{\secref}[1]{Section~\ref{#1}}

\newcommand{\tabref}[1]{Table~\ref{#1}}

\newcommand{\listing}[1]{Listing~\ref{#1}}
\def \ULYSSES{\sc ULYSSES}
\usepackage{pgf}

\biboptions{sort&compress} 

\begin{document}

\AddToShipoutPictureBG*{%
  \AtPageUpperLeft{%
    \hspace{0.95\paperwidth}%
    \raisebox{-\baselineskip}{%
      \makebox[1pt][r]{IPPP/23/02}
      }
      }
      }%

\begin{frontmatter}
\title{{\ULYSSES}, Universal LeptogeneSiS Equation Solver:\\
version 2}

\author[a,b]{A.~Granelli}
\author[c]{C.~Leslie}
\author[c]{Y. F.~Perez-Gonzalez}
\author[d]{H.~Schulz}
\author[e]{B.~Shuve}
\author[c]{J.~Turner}
\author[c]{R.~Walker}

\address[a]{Dipartimento di Fisica e Astronomia, Università di Bologna, via Irnerio 46, 40126, Bologna, Italy}
\address[b]{INFN, Sezione di Bologna, viale Berti Pichat 6/2, 40127, Bologna, Italy}
\address[c]{Institute for Particle Physics Phenomenology, Durham University, DH1 3LF, Durham, UK}
\address[d]{Arrival Ltd, Beaumont House, Kensington Village, W14 8TS, London, UK}
\address[e]{Harvey Mudd College, 301 Platt Blvd., Claremont, CA 91711, United States}

\journal{Computer Physics Communications}

\begin{abstract}
{\ULYSSES} is a Python package that calculates the baryon asymmetry 
produced from leptogenesis in the context of a type-I seesaw mechanism. 
In this release, the new features include code which solves the Boltzmann equations for low-scale leptogenesis;
the complete Boltzmann equations for thermal leptogenesis applying proper quantum statistics without assuming kinetic equilibrium of the right-handed neutrinos; and, primordial black hole-induced leptogenesis.
{\ULYSSES} version 2 has the added functionality of a pre-provided script for a two-dimensional grid scan of the parameter space. As before, the emphasis of the code is on
user flexibility, rapid evaluation and is publicly available at \url{https://github.com/earlyuniverse/ulysses}.
\end{abstract}
\end{frontmatter}

\section{Introduction}\label{sec:overview}
\noindent Since its initial proposal \cite{Fukugita:1986hr}, leptogenesis has been one of the most well-studied mechanisms to explain matter-antimatter asymmetry. An appealing additional aspect of this mechanism is its connection with neutrino masses and mixing. 
{\ULYSSES} \cite{Granelli:2020pim} is a \emph{Python} package that solves the semi-classical Boltzmann equations (BEs) for leptogenesis in the context of a type-I seesaw mechanism \cite{Minkowski:1977sc, Yanagida:1979as, GellMann:1980vs, Glashow:1979nm, Mohapatra:1979ia}. {\ULYSSES} version 1, presented in Ref.~\cite{Granelli:2020pim}, provided code for solving the momentum-averaged BEs relevant to leptogenesis based on the out-of-equilibrium decays of right-handed neutrinos (RHNs) for both resonant and non-resonant regimes. In addition, effects such as lepton flavour, scatterings and spectator processes are available if the user wishes to apply them.  
In this updated version, we provide additional BEs codes which solve the ``complete'' set of thermal leptogenesis BEs \cite{Hahn-Woernle:2009jyb} that properly accounts for quantum statistics and does not assume kinetic equilibrium for the RHNs. Furthermore, we provide state-of-the-art BEs for low-scale (also known as ARS) leptogenesis via oscillations \cite{Akhmedov:1998qx,Asaka:2005pn} and primordial black hole-induced thermal leptogenesis \cite{Perez-Gonzalez:2020vnz,Bernal:2022pue}.

For a given point in the model parameter space, {\ULYSSES} calculates the final baryon asymmetry (given in terms of the baryon-to-photon ratio, $\eta_{B}$, the baryonic yield, $Y_B$, and the baryonic density parameter, $\Omega_B h^2$) and plots the lepton asymmetry number density as a function of 
the evolution parameter. For the user who wishes to undertake a multi-dimensional exploration of the parameter space, we have provided instructions on how to use {\sc Multinest} \cite{Feroz:2008xx} in combination with {\ULYSSES} in the manual of {\ULYSSES} version 1 \cite{Granelli:2020pim}. 
{\ULYSSES} is designed modularly, separating the physics of the baryon asymmetry production
from the parameter space exploration. 
In this paper, we will not recapitulate on the BEs provided in version 1, but present the new features along with the basics of {\ULYSSES} installation and functionality. {\ULYSSES} version 2 applies all of the same conventions (Yukawa matrix parametrisation, Higgs vacuum expectation value, normalisation of number densities) as version 1, and we refer the reader to the previous version for discussion on such matters. 
Further, we refrain from discussing the different regimes and subtleties of the leptogenesis mechanism and instead refer the reader to recent reviews (see, \eg,  Refs.~\cite{Bodeker:2020ghk,Asadi:2022njl} and references therein) on various aspects of thermal, resonant and low-scale leptogenesis.

The paper is organised as follows: in \secref{sec:plugins}, we describe the new pre-provided BEs and follow in \secref{sec:install} with installation instructions.
In \secref{sec:usage}, we detail the usage of the {\ULYSSES} version 2: namely, we write how to input the model parameters, introduce the new functionality, (that is, the two-dimensional grid scan of the parameter space), and give specific examples on how to call the new model files and their related example parameter cards. Finally, we make concluding remarks in \secref{sec:conclusions}.

\section{Newly Added Built-in Boltzmann Equations}\label{sec:plugins}
\noindent In this section, we list and  discuss the \emph{newly added}  BEs that are shipped with
 {\ULYSSES} version 2. We note that version 2 contains the same BEs as version 1, detailed in the previous manual version.
 In \secref{sec:TL} and \secref{sec:ARS}, which discuss the BEs for thermal leptogenesis and leptogenesis via oscillations, standard cosmology is assumed. However, in \secref{sec:PBH}, non-standard cosmology, including a population of primordial black holes, is included in the evolution of the lepton asymmetry.
 \subsection{Thermal Leptogenesis}\label{sec:TL}
\noindent The majority of papers in the literature apply the following two assumptions: 
\begin{enumerate}
\item  The phase space distribution functions, $f_i$, of the particles species involved in leptogenesis, $i = N, \Phi,\,l$, when in thermal equilibrium, are approximated by a Maxwell-Boltzmann distribution, $f_{i}^{\mathrm{eq}}=e^{-E_{i} / \mathrm{T}}$. Within the Maxwell-Boltzmann statistics, it is also a good approximation to neglect the quantum Pauli-blocking (Bose-enhancement) factors for fermions (bosons), \ie,~$1-f^\text{eq}_i\simeq 1$ ($1+f^\text{eq}_i \simeq 1$). \\

\item  The RHNs are in kinetic equilibrium, $f_{\mathrm{N}}/f_{\mathrm{N}}^{\mathrm{eq}} \approx n_{\mathrm{N}}/n_{\mathrm{N}}^{\mathrm{eq}}$, where $n_N^\text{(eq)}$ is the (equilibrium) number density of RHNs.
\end{enumerate}
\begin{table}[t]
\centering
\begin{tabular}{c|c|c|}
\cline{2-3}
    & \textbf{Maxwell Boltzmann statistics} & {\textbf{RHN Kinetic Equilibrium}} \\ \hline
\multicolumn{1}{|c|}{{{Case 1}}} &  {\checkmark}  &  {\checkmark}  \\ \hline
\multicolumn{1}{|c|}{{{Case 2}}} &  {\checkmark}  & {$\times$}  \\ \hline
\multicolumn{1}{|c|}{{{Case 3}}} & {$\times$}  & {\checkmark} \\ \hline
\multicolumn{1}{|c|}{{{Case 4}}} & {$\times$}  &   {$\times$}  \\ \hline
\end{tabular}
\caption{The combinations of assumptions used for each case where ${\checkmark}$ denotes when the assumption is applied and {$\times$} when it is neglected.}
\label{tab:cases}
\end{table}
The effects of dropping these assumptions were studied in detail in Ref.~\cite{Hahn-Woernle:2009jyb}.

In \tabref{tab:cases}, we show four cases where these assumptions are applied or neglected. 
For {Case 1} -- {Case 4} only the decays and inverse decays of the RHN are included; however, Ref.~\cite{Hahn-Woernle:2009jyb} outlines how to include the 
effect of scattering and we relegate the inclusion of such an effect for a  future  {\ULYSSES} version. In all cases, the initial RHN and lepton asymmetry abundance ($N_{B-L}$) is set using an array $\texttt{y0}$ in the model files, and the default initial abundance is zero-valued.

\noindent{Case 1} is simply the standard momentum-averaged, one decaying RHN, single-flavoured BEs in which
Maxwell-Boltzmann statistics {and RHN kinetic equilibrium are} assumed. This is shipped with {\ULYSSES} under the code name \texttt{etaB1BE1F.py} and shortcut name \texttt{1BE1F} (already included in the first version of the code \cite{Granelli:2020pim}).
The BEs for this simple case are 
\begin{subequations}
\begin{align}
    \frac{d N_{\mathrm{N}}}{d z} &= -D\left(N_{\mathrm{N}}-N_{\mathrm{N}}^{\mathrm{eq}}\right)\,,\\
    \frac{d N_{B-L}}{d z} &= \epsilon D\left(N_{\mathrm{N}}-N_{\mathrm{N}}^{\mathrm{eq}}\right)-W N_{B-L}\,,
    \end{align}
\end{subequations}
where $z \equiv M_1/T$, $M_1$ is the mass of the lightest RHN and $T$ is the temperature of the plasma; the quantities $N_N^\text{(eq)}$
and $N_{B-L}$ are respectively the number of RHNs (when in thermal equilibrium) and $B-L$ asymmetry in a comoving volume normalised to contain one photon when $z\ll 1$ \footnote{Considering a Bose-Einstein distribution for photons, our choice of normalisation means that $g_\gamma \zeta(3) T^3a^3/\pi^2 = 1$, where $a^3$ is the comoving volume, $g_\gamma = 2$ are the photon degrees of freedom and $\zeta$ is the Riemann zeta function with $\zeta(3)\simeq 1.20$. Adopting this normalisation at $z \ll 1$ within the Maxwell-Boltzmann statistics for RHNs and leptons leads to $N_{N,\,l}^\text{eq} = 1/\zeta(3)$, while a more accurate Fermi-Dirac distribution would give $N_{N,\,l}^\text{eq} = 3/4$ (this discrepancy was also noted in Ref.~\cite{Hahn-Woernle:2009jyb}). 
As in the first version of the code \cite{Granelli:2020pim}, we adopt the analytical approximation $N_{N,\,l}^\text{eq}(z) \simeq (3/8) z^2K_2(z)$, $K_n(z)$ being the modified $n^\text{th}$ Bessel functions of the second kind, to match the Maxwell-Boltzmann statistics at $z\gtrsim 1$ and get the correct normalisation condition when $z\ll 1$.}; $D$ and $W$ are respectively the decay and washout parameters (see, \eg, Refs.~\cite{Hahn-Woernle:2009jyb, BUCHMULLER2005305} for specific expressions); and $\epsilon$ is the CP-asymmetry parameter.

\noindent{Case 2} solves the following coupled differential system (Case D2 of Ref.~\cite{Hahn-Woernle:2009jyb}):
\begin{subequations}
\begin{align}
  \frac{\partial f_{\mathrm{N}}}{\partial z} & =\frac{z^{2} K}{\mathcal{E}_{\mathrm{N}}}\left(e^{-\mathcal{E}_{\mathrm{N}}}-f_{\mathrm{N}}\right)\,,\\
    \frac{d N_{B-L}}{d z} &=-\frac{z^{2} K}{4\zeta(3)} \int_0^\infty {d}y_l \int_{\left | \frac{4 y_{l}^{2}-z^2}{4 y_{l}} \right |}^{\infty} {d} y_{\mathrm{N}} \frac{y_{\mathrm{N}}}{\mathcal{E}_{\mathrm{N}}} \left[\frac{4}{3} f_{\mathrm{N}}^{\mathrm{eq}}  N_\lminus-2 \epsilon\left(f_{\mathrm{N}}-f_{\mathrm{N}}^{\mathrm{eq}}\right)\right]\,,
\end{align}
\end{subequations}
where $K$ is the decay parameter, $\mathcal{E}_{\mathrm{N}}$ is the RHN energy normalised to temperature, $y_{\mathrm{i}}$ is the modulus of the three-momentum of species $i$ normalised to temperature, and $f_{\mathrm{N}}^{\mathrm{eq}}=e^{-\mathcal{E}_{\mathrm{N}}}$.
As the RHN phase space distribution is not initially integrated over energy, there are two integrals in the second coupled differential equation: the first is over the 
RHN three-momentum normalised to temperature ($y_\mathrm{N}$); the second is over the three-momentum of the lepton $l$ normalised to temperature ($y_l$). Since there are two integrations per time step, Case 2 is computationally more expensive than Case 1. We note that $f_N$ is calculated on a grid of $y_N$ and then interpolated to perform this integration.  The BE code containing Case 2 is \texttt{etab1BE1F$\_$Case2.py} and the shortcut name is \texttt{1BE1F$\_$Case2}.

\begin{figure}[t!]
    \centering
    \includegraphics[width=.9\textwidth]{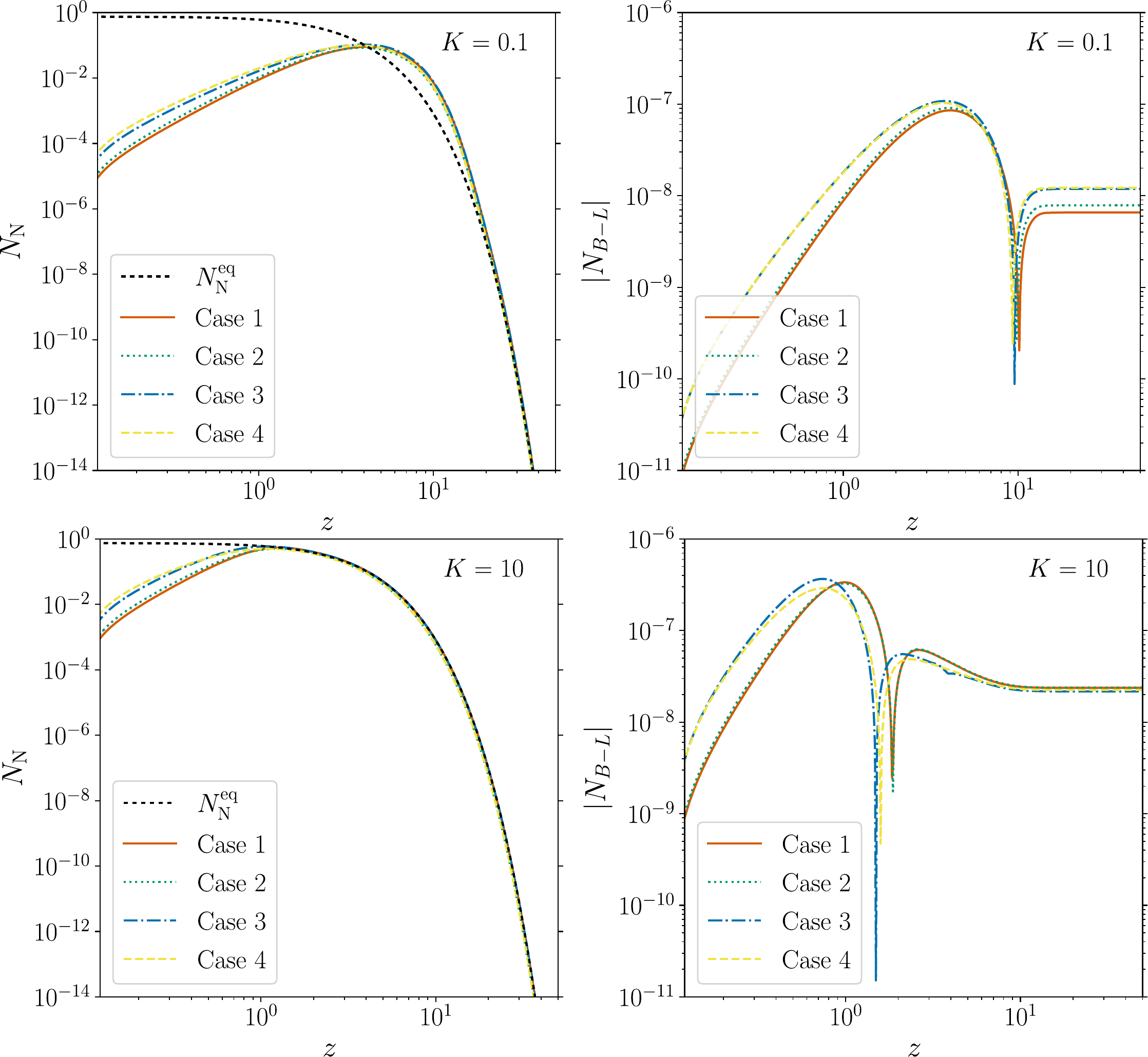}
    \caption{Time evolution of the comoving RHN number density (left) and the absolute value of the lepton asymmetry (right)
 assuming the decay  and CP asymmetry parameter have values $K=10\, (K=0.1)$ and $\epsilon=10^{-6}$ respectively for the top (bottom) panels.
Solid/red line denotes Case 1, 
dotted  green Case 2, dotted-dashed blue Case 3 and dashed yellow Case 4.}
    \label{fig:Keq10plot}
\end{figure}
\noindent{Case 3} solves the following coupled differential system (Case D3 of Ref.~\cite{Hahn-Woernle:2009jyb}):
\begin{subequations}
\begin{align}
\frac{d N_\mathrm{N}}{d z} &=  \frac{K}{{\color{red}} K_2(z)} \left(N_{\mathrm{N}} - N_{\mathrm{N}}^{\mathrm{eq}}\right) \int_0^\infty {d} y_{\mathrm{N}} \; \frac{y_{\mathrm{N}}}{\mathcal{E}_{\mathrm{N}}}  \frac{1}{\left(e^{\mathcal{E}_{\mathrm{N}}}+1\right)} \;\log \left \{\frac{\sinh\left[\left(\mathcal{E}_{\mathrm{N}} - y_{\mathrm{N}}\right)/2\right]}{\sinh\left(\left(\mathcal{E}_{\mathrm{N}} + y_{\mathrm{N}}\right)/2\right]}\right\}\,,\\
  \frac{d N_{B-L}}{d z} &=-\frac{z^{2} K}{4\zeta(3)} \int_{0}^{\infty} {d} y_{l} \int_{\left | \frac{4 y_{l}^{2}-z^2}{4 y_{l}} \right |}^{\infty} {d} y_{\mathrm{N}} \frac{y_{\mathrm{N}}}{\mathcal{E}_{\mathrm{N}}} \Bigg[\left(f_{\Phi}+\frac{N_{\mathrm{N}}}{N_{\mathrm{N}}^{\mathrm{eq}}} f_{\mathrm{N}}^{\mathrm{eq}}\right)\left(\frac{4}{3} N_{B-L}+2 \epsilon\right) f_{l}^{\mathrm{eq}}+\nonumber\\&\hspace{23em}-
  2 \epsilon \frac{N_{\mathrm{N}}}{N_{\mathrm{N}}^{\mathrm{eq}}} f_{\mathrm{N}}^{\mathrm{eq}}\left(1+f_{\Phi}\right)\Bigg].
\end{align}
\end{subequations}
In the set of equations given above, quantum statistics are applied, but kinetic equilibrium for the RHN is assumed.
The integral on the right-hand side of the first equation has no simple analytic form, and it
is necessary to perform the integration numerically. The BE code containing Case 3 is \texttt{etab1BE1F$\_$Case3.py}, and the shortcut name is \texttt{1BE1F$\_$Case3}.

\noindent{Case 4} solves the following coupled differential system (Case D4 of Ref.~\cite{Hahn-Woernle:2009jyb}):
\begin{subequations}
\begin{align}
\frac{\partial f_{\mathrm{N}}}{\partial z} & = \frac{z^2K}{\mathcal{E}_{\mathrm{N}} y_{\mathrm{N}}} \frac{f_{\mathrm{N}}(1+e^{\mathcal{E}_{\mathrm{N}}})-1}{\left(e^{\mathcal{E}_{\mathrm{N}}}+1\right)}
\log \left \{\frac{\sinh\left[\left(\mathcal{E}_{\mathrm{N}} - y_{\mathrm{N}}\right)/2\right]}{\sinh\left(\left(\mathcal{E}_{\mathrm{N}} + y_{\mathrm{N}}\right)/2\right]}\right\}\,,\\
    \frac{d N_{B-L}}{d z} & =-\frac{z^{2} K}{4\zeta(3)} \int_{0}^{\infty} {d} y_{l}  \int_{\left | \frac{4 y_{l}^{2}-z^2}{4 y_{l}} \right |}^{\infty} {d} y_{\mathrm{N}} \frac{y_{\mathrm{N}}}{\mathcal{E}_{\mathrm{N}}}\Bigg[\left(f_{\Phi}+f_{\mathrm{N}}\right)\left(\frac{4}{3} N_{B-L}+2 \epsilon\right) f_{l}^{\mathrm{eq}}+\nonumber\\&\hspace{23em}-2 \epsilon f_{\mathrm{N}}\left(1+f_{\Phi}\right)\Bigg]\,,
    \end{align}
\end{subequations}
where the quantum statistics have been correctly accounted for, and the kinetic equilibrium of the RHN is not assumed. The BE code containing {Case 4} is \texttt{etab1BE1F$\_$Case4.py}, and the shortcut name is \texttt{1BE1F$\_$Case4}.
In \figref{fig:Keq10plot}, we show the solutions of Cases 1 to 4 for weak ($K=0.1$) and strong ($K=10$)
washout. It is well known that, in the strong washout regime ($K\gg1$), the set of BEs with the assumption of kinetic equilibrium and Maxwell-Boltzmann statistics provides a solution that is quantitatively similar to that of the complete BEs \cite{Hahn-Woernle:2009jyb}. In the weak washout regime ($K<1$) the difference can be up to a factor of $\sim 2$, mainly due to using the correct Bose-Einstein equilibrium distribution function for the Higgs boson, substantially enlarging the phase space available for the inverse decay process.

 To solve Case 2 to 4, \texttt{solve}$\_$\texttt{ivp} is used with Runge-Kutta order 5. Although this method was computationally more expensive than the third-order Runge-Kutta method, it provides much more stable and accurate results. Finally, while the integrations in $y_N$ and $y_{l}$ formally have infinity as an upper integration boundary, we found that an upper limit of $300$ is more than sufficient for larger values, and the distributions function is effectively zero-valued. 

\subsection{Leptogenesis with GeV-Scale Right-Handed Neutrinos}\label{sec:ARS}
\noindent When the RHN masses are at the GeV-scale, the na\"ive seesaw mechanism predicts that they have Yukawa couplings to the SM leptons and Higgs of the order of $\sim 10^{-7}$ (see, \eg, Ref.~\cite{Ghiglieri:2017dee}). Consequently, the RHNs are expected to be out of equilibrium in the early Universe, allowing a lepton asymmetry to be generated during their production and approach to equilibrium (``freeze-in'') rather than exclusively during their departure from it (``freeze-out''). This is the Akhmedov-Rubakov-Smirnov (ARS) mechanism for leptogenesis via oscillations of RHNs \cite{Akhmedov:1998qx,Asaka:2005pn}. This mechanism, in which the observed baryon asymmetry can be generated prior to the electroweak phase transition from the dynamics of GeV-scale RHNs, has been extensively studied in recent years (see, \eg, Refs.~\cite{Ghiglieri:2017dee, hepph0605047,Asaka:2011wq,Canetti:2012kh,Shuve:2014zua,1508.03676,1606.06690,Hernandez:2016kel,Asaka:2017rdj,Ghiglieri:2017csp,Drewes:2017zyw,Abada:2018oly}) and has recently received further attention because of its compatibility with couplings of the RHNs to the charged and neutral SM currents that could be accessible at future accelerator and collider experiments \cite{Klaric:2020phc, Klaric:2021cpi, Drewes:2021nqr, Abada:2018oly} (for a recent review, see also Ref.~\cite{Chun:2017spz}), as well as to current and upcoming experiments on charged lepton flavour violating processes involving muons \cite{Granelli:2022eru, Calderon:2022alb, Hernandez:2022ivz}.

The asymmetry in ARS leptogenesis results from the interplay of CP-violating phases in the RHN Yukawa couplings and oscillation phases among linear combinations of RHN mass eigenstates. Consequently, the semi-classical BE approach to thermal leptogenesis is inadequate as it does not keep track of coherences among states. Instead, the RHN abundances should be modelled as a set of density matrices, and a set of quantum kinetic equations (QKEs) must be solved for the simultaneous evolution of the RHN abundances and lepton flavour asymmetries. SM flavour effects are essential:~in the minimal ARS scenario, the initial lepton asymmetry sums to zero over all flavours, and a net baryon asymmetry  results only due to subsequent flavour-dependent washout of each flavour asymmetry \cite{Akhmedov:1998qx,Asaka:2005pn}.

The QKEs for the RHN density matrices consist of two types of terms:~oscillation terms, consisting of commutators of the density matrices with the mass terms in the Hamiltonian (originating  from both tree-level  and finite-temperature contributions), which account for the oscillation phases; and collision terms, which produce/destroy specific linear combinations of RHN mass eigenstates and also lead to decoherence. Some collision terms are independent of the chemical potentials in SM leptons and allow for the generation of initial lepton flavour asymmetries. In contrast, other terms depend on the lepton chemical potentials and account for back-reactions and washout of the lepton flavour asymmetries.

Since the oscillation phases depend on the momentum of the particular RHN state involved, one must in principle set up QKEs for each RHN momentum mode and separately solve for the asymmetry generated by each mode. This is very computationally intensive and impractical for large-scale studies. Therefore, it is more feasible to instead perform an average over RHN momentum in the oscillation and collision terms and derive QKEs for the momentum-averaged RHN density matrices. Dedicated studies comparing the proper and momentum-averaged treatments typically show agreement up to $\mathcal{O}(1)$ factors (although for individual points the discrepancy can be higher) \cite{Ghiglieri:2017csp,Ghiglieri:2018wbs}; this level of precision is sufficient for most studies and so we adopt this procedure \footnote{Momentum-dependent asymmetries were also computed in a related model where the oscillating states were produced in the decay of a heavy scalar, with similar conclusions about the accuracy of the momentum averaging procedure \cite{Shuve:2020evk,Berman:2022oht}.}. 

We implement in {\ULYSSES} the momentum-averaged QKEs relevant to ARS leptogenesis with two RHNs that are quasi-degenerate in mass with masses $M_1\simeq M_2$, including both lepton-number-conserving (LNC) and lepton-number-violating (LNV) terms. The LNV terms are proportional to the RHN Majorana masses and are consequently suppressed by $M_{1,\,2}^2/T^2$ relative to the LNV terms for $T\gg M_{1,\,2}$, but they can be important for RHN masses close to or above the electroweak scale. The QKEs for ARS leptogenesis have been derived with varying levels of refinement in Refs.~\cite{Akhmedov:1998qx,Asaka:2005pn,Asaka:2011wq,Canetti:2012kh,Hernandez:2016kel,Hambye:2017elz,Ghiglieri:2017dee,Ghiglieri:2017csp,Eijima:2018qke,Abada:2018oly,Klaric:2020phc,Klaric:2021cpi,Drewes:2021nqr,Hernandez:2022ivz}. We adopt a notation similar to that of Ref.~\cite{Hernandez:2022ivz} (see also Refs.~\cite{Hernandez:2016kel,Abada:2018oly}), which is physically transparent and correct in the limit of relativistic RHNs, while also admitting a relatively simple approximation for obtaining approximate results beyond the leading expansion in $\mathcal{O}(M_{1,\,2}^2/T^2)$. The QKEs are written in terms of $(R_N)_{IJ}\equiv (n_N)_{IJ}/n^{\rm eq}_N$ and $(R_{\overline N})_{IJ}$, the RHN and anti-RHN density matrices normalised to the equilibrium abundance, as well as the lepton chemical potentials normalised to the temperature, $\mu_{B/3-L_\alpha}$ and $\mu_\alpha$. More specifically, the quantity $\mu_\alpha$ is the reduced chemical potential in the lepton doublet of flavour $\alpha$, while $\mu_{B/3-L_\alpha}$ parameterises the asymmetry in the anomaly-free charge $B/3-L_\alpha$ that is conserved by SM interactions (not summed over gauge degrees of freedom in the lepton doublet). Spectator effects relate $\mu_{B/3-L_\alpha}$ and $\mu_\alpha$ according to \cite{Abada:2018oly}
\begin{eqnarray}
\mu_\alpha &=& 2\sum_\beta\,\chi_{\alpha\beta}\, \mu_{B/3-L_\beta}\,,\\
\left(\chi_{\alpha\beta}\right)&=& -\frac{1}{711}\left(\begin{array}{ccc}
257 & 20 & 20 \\
20 & 257 & 20 \\
20 & 20 & 257 \end{array}\right)\,.
\end{eqnarray}
The relation to the $B-L$ asymmetry yield (after summing over lepton gauge degrees of freedom) is 
\begin{eqnarray}
Y_{B-L} &=& \sum_\alpha\frac{15}{2\pi^2g_{*,\,s}}\,\mu_{B/3-L_\alpha}\,,
\end{eqnarray}
where $g_{*,\,s}$ is the number of entropic degrees of freedom, and the ratio between the baryon and $B-L$ asymmetries is the usual factor of $28/79$. 

The explicit form of the QKEs we implement is in dimensionless form,
%
\begin{subequations}
\begin{align}
\frac{T_{\rm ew}}{M_0}\frac{dR_N}{dz} &= -i\frac{z}{T_{\rm ew}}\left[\langle H\rangle,R_N\right] - \frac{1}{2}\frac{\langle\gamma_N^{(0)}\rangle}{T}\left\{F^\dagger F,R_N-1\right\} + \frac{\langle\gamma_N^{(1)}\rangle}{T}F^\dagger \mu F\nonumber\\
&{} -\frac{1}{2}\frac{\langle\gamma_N^{(2)}\rangle}{T}\left\{F^\dagger\mu F,R_N\right\} - \frac{z^2}{2T_{\rm ew}^2}\frac{\langle S_N^{(0)}\rangle}{T}\left\{ MF^{\rm T}F^*M,R_N-1\right\}\nonumber\\
&{} -\frac{z^2}{T_{\rm ew}^2}\frac{\langle S_N^{(1)}\rangle}{T}\,M F^{\rm T}\mu F^* M + \frac{z^2}{2T_{\rm ew}^2}\frac{\langle S_N^{(2)}\rangle}{T}\left\{MF^{\rm T}\mu F^* M,R_N\right\}\nonumber\\
&-\frac{T_{\rm ew}}{M_0}\frac{R_N}{Y_N^{\rm eq}}\frac{dY_N^{\rm eq}}{dz},\\
\frac{2\pi^2}{9\zeta(3)}\frac{T_{\rm ew}}{M_0}\frac{d\mu_{B/3-L_\alpha}}{dz} &= -\frac{1}{2}\frac{\langle\gamma_N^{(0)}\rangle}{T}\left(FR_NF^\dagger-F^* R_{\overline N} F^{\rm T}\right)_{\alpha\alpha} + \frac{\langle\gamma_N^{(1)}\rangle}{T}\left(FF^\dagger\right)_{\alpha\alpha}\mu_\alpha\nonumber\\
&{}-\frac{1}{2}\frac{\langle\gamma_N^{(2)}\rangle}{T}\left(FR_NF^\dagger+F^*R_{\overline N}F^{\rm T}\right)_{\alpha\alpha}\mu_\alpha\nonumber\\
&{}+ \frac{z^2}{2T_{\rm ew}^2}\frac{\langle S_N^{(0)}\rangle}{T}\left(F^*MR_NMF^{\rm T}-FMR_{\overline N}MF^\dagger\right)_{\alpha\alpha} \nonumber\\&+ \frac{z^2}{T_{\rm ew}^2}\frac{\langle S_N^{(1)}\rangle}{T}\left(FM^2F^\dagger\right)_{\alpha\alpha}\mu_\alpha \nonumber\\
&{}-\frac{z^2}{2T_{\rm ew}^2}\frac{\langle S_N^{(2)}\rangle}{T}\left(FMR_{\overline N}MF^\dagger+F^*MR_NMF^{\rm T}\right)_{\alpha\alpha}\mu_\alpha\,,
 \end{align}
 \end{subequations}
 %
 and the $R_{\overline{N}}$ equation is found by taking $R_N\to R_{\overline N}$, $\mu \to-\mu$ and $F\to F^*$ in the QKE for $R_N$. We define $z\equiv T_{\rm ew}/T$, $M_0\equiv M_{\rm Pl}/(1.66\sqrt{g_{*,\,s}})\approx7.1\times10^{17}$ GeV (so that the Hubble rate is $H(T) = T^2/M_0$), $\mu$ as the diagonal matrix of lepton doublet chemical potentials $\mu_\alpha$, $M$ as the diagonal matrix of RHN Majorana masses, and $T_{\rm ew}\approx 131.7$ GeV as the temperature of sphaleron decoupling \footnote{We warn the reader on the different definition we have adopted in this context for the time variable $z = T_\text{ew}/T$, while, in the other modules, we have used $z = M_1/T$. We adopt this convention because the asymmetry depends predominantly on the squared mass splitting $M_2^2-M_1^2$ rather than on the absolute masses $M_{1,2}$, and consequently it is more convenient to normalize to $T_{\rm ew}$.}. 

 The thermally averaged Hamiltonian, including both tree-level masses and the effective potential induced by the medium in the high-temperature expansion, is \cite{Hernandez:2016kel,Abada:2018oly}
\begin{equation}
\langle H\rangle_{IJ} = \frac{\pi^2 z}{36\zeta(3)T_{\rm ew}}M^2_{IJ} + \frac{\pi^2 T_{\rm ew}}{144\zeta(3)z} (F^\dagger F)_{IJ}\,,
\end{equation}
%
with $I,\,J = 1,\,2$ labelling the zero-temperature RHN mass eigenstates. Given that a multiple of the identity matrix can be added to the Hamiltonian without changing the dynamics (it only leads to an irrelevant overall phase), we subtract the overall mass scale in $M^2$, allowing us to replace $M^2$ with a diagonal matrix with entries $(0,M_2^2-M_1^2)$. This makes the computations faster as we do not need to keep track of the irrelevant phase given by the overall mass scale.
 
 The reaction rates (stripped of coupling constants and powers of $M_{1,2}$) are labelled by $\langle \gamma_N^{(j)}\rangle$ for LNC rates and $\langle S_N^{(j)}\rangle$, with $j = 0,\,1,\,2$, for LNV rates:~note that the LNV processes are evaluated in the limit of highly relativistic RHNs and are accompanied in the QKEs with multiplicative powers of $M_{1,\,2}^2/T^2$, as expected. We perform a thermal average over the momentum-dependent, temperature-normalised rates from Ref.~\cite{Ghiglieri:2017csp} \footnote{The rates from Ref.~\cite{Ghiglieri:2017csp} are provided in tabular form at \href{http://www.laine.itp.unibe.ch/leptogenesis/}{\ttfamily http://www.laine.itp.unibe.ch/leptogenesis/}. We thank Stefan Sandner for pointing us towards this electronic database and the public AMIQS code at \url{https://github.com/stefanmarinus/amiqs} based on Ref.~\cite{Hernandez:2022ivz}. We have verified that our momentum-averaged reaction rates agree with those from Ref.~\cite{Hernandez:2022ivz}.}. The rates labelled with $j=0$ are independent of the lepton chemical potentials, while those corresponding to $j=1,2$ depend on lepton chemical potential, with the former being independent of RHN abundances and the latter depending on both lepton chemical potentials and RHN abundances. The ratios $\langle \gamma^{(j)}\rangle/T$ and $\langle S^{(j)}\rangle/T$ are largely independent of the temperature for $T\gg T_{\rm ew}$, and so, by default, {\ULYSSES} fixes these ratios to their values at $T=10^3$ GeV. However, we have provided an option for the user to include the temperature dependence of these ratios and to take into account higher-order non-relativistic contributions to the LNV terms as in \cite{Hernandez:2022ivz} \footnote{Non-relativistic corrections as in Ref.~\cite{Hernandez:2022ivz} for the LNC rates are negligible for mass scales below 100 GeV (see, \eg, left panel of Fig.~3 of Ref.~\cite{Hernandez:2022ivz}), but  we still have provided tables for interpolation for the user who wishes to include such corrections.} (further information is provided in \secref{sec:usage}).

 The QKEs are solved from a user-specified initial time $z_0$ to a final time $z=1$, which is the time of sphaleron decoupling. The user can also specify the initial RHN density matrix at $z=z_0$ in the model file. 
 Two typical choices of initial conditions are $R_N=0$ and $R_N=1$, corresponding, respectively, to vanishing or thermal initial conditions; in the latter case, the asymmetry cannot be generated in the freeze-in regime, as there is no initial net production of RHNs given that they start in equilibrium, and the asymmetry only results from 
the freeze-out mechanism. The default abundance is set to be $R_N=0$, but can easily be adjusted by the user by changing the \texttt{y0} array.
 Other choices of the initial condition can substantially enhance or decrease the resulting final asymmetry \cite{Shuve:2020evk, Asaka:2017rdj}.
 
 Typically, the QKEs are a stiff system of differential equations because multiple time scales exist corresponding to oscillation and equilibration of the RHNs. Furthermore, the oscillation frequency increases at later times; this can present a challenge to the numerical integration of the QKEs, particularly for earlier onsets of oscillations (equivalent to larger values of $\Delta M_{21}^2 = M_2^2-M_1^2$). The determination of the asymmetry is simplified by the fact that the generation of lepton flavour asymmetries is suppressed after the onset of rapid oscillations because the positive and negative contributions to the lepton asymmetry average to zero at this point \cite{Asaka:2005pn}, and, consequently, there is little value in tracking the phase information of RHNs past this point. The onset of RHN oscillations occurs around the dimensionless time given by
 \begin{equation}
 z_{\mathrm{osc}}\equiv\left(\frac{12 T_{\mathrm{ew}}^3}{\Delta M_{21}^2 M_0}\right)^{1 / 3}.
\end{equation}
 For $z_{\mathrm{osc}}>0.1$, the epoch of rapid oscillations is sufficiently close to the electroweak time $z=1$ that we solve the full set of QKEs including all oscillations.
 
 For $z_{\rm osc}<0.1$, we offer the user a ``stitching" option to truncate the generation of the flavour asymmetries at a  dimensionless time $z_{\rm cut}> z_{\rm osc}$:~the full set of QKEs are solved from $z_0$ to $z_{\rm cut}$, and the solutions at $z_{\rm cut}$ are used as the initial conditions for a new set of QKEs with the off-diagonal components of $R_N$ and $R_{\overline N}$ set to 0 (in the epoch of rapid oscillations the off-diagonal terms all average to zero). This latter set of QKEs is then solved to the final time $z=1$. The default value of $z_{\rm cut}$ is set to 1, equivalent to solving the full QKEs for the entire time interval relevant for generating the baryon asymmetry. However, the user can specify an alternative value of the stitching time by changing the parameter \texttt{zcut}. 
This approach allows for rapid integration of the equations for tracking washout effects if the subsequent generation of the asymmetry beyond $z_{\rm cut}$ is known to be small. However, we warn the user that this stitching functionality should only be applied if the necessary condition is met and that an appropriate value of $z_{\rm cut}$ has been selected:~for instance, the final result should not change under modest adjustments to $z_{\rm cut}$, meaning that the full solution has only been truncated when the off-diagonal terms average to zero.
Additionally, the user should validate their procedure for a few parameter points by comparing their  solution using the stitching option to the full solutions with $z_{\rm cut}=1$. 

 \emph{Validity and limitations:}~The QKEs and reaction rates implemented in {\ULYSSES} are expected to be valid for GeV-scale RHNs ($M_{1,\,2}\ll T_{\rm ew}$) and may give reasonable estimates for somewhat larger masses. However, increasing the RHN masses makes the LNV terms more relevant, and the LNV rates exhibit a pronounced temperature dependence in the vicinity of the electroweak crossover. While our treatment of the QKEs has an option to include this temperature dependence and, to some extent, the non-relativistic contributions for the LNV terms as in Ref.~\cite{Hernandez:2022ivz}, other effects are not currently treated in {\ULYSSES}, such as (more precise) higher-order corrections to the energy-momentum relation \cite{Klaric:2021cpi,Hernandez:2022ivz} and a non-instantaneous treatment of sphaleron decoupling \cite{Eijima:2017cxr}. These effects are more important for larger RHN masses and/or larger coupling (\emph{i.e.,} in the strong washout limit where all the RHNs and SM leptons of each flavour come into equilibrium), and can affect the baryon asymmetry by up to an order of magnitude. Caution is merited when using these results for RHN masses approaching or exceeding 100 GeV and/or in the strong washout limit.
\subsection{Leptogenesis from Primordial Black Hole Evaporation}\label{sec:PBH}
\noindent After discovering Gravitational Waves from Black Hole mergers, analysing the properties and phenomenological effects in Astrophysics and Cosmology of Black Holes has seen a renewed interest.
One interesting effect is understanding the possible consequences of Primordial Black Hole (PBH) evaporation on different particle phenomena in the Early Universe.
Leptogenesis indeed can be affected if there existed a non-negligible population of evaporating PBHs~\cite{Perez-Gonzalez:2020vnz, Bernal:2022pue}.
Since RHNs would be among the particles emitted by the PBHs, their CP-violating decays could produce more baryon asymmetry than those created in the primordial plasma.
Moreover, depending on when the evaporation occurs, the washout effects could be out of equilibrium so that this new population of RHNs would not erase the pre-existing asymmetry.
Additionally, if the PBHs had large initial masses, the effect would be different; they inject a large amount of entropy that could dilute the previous lepton asymmetry in the plasma.
In order to determine the baryon asymmetry correctly, we need to track in detail the evolution of all Universe components: radiation and PBH energy densities, together with the RHN and $B-L$ number densities.

Kerr PBHs are characterised by their mass $\Mbh$ and spin parameter $a_* \equiv J/(G^2\Mbh^2)\in [0, 1)$, with $J$ being the BH angular momentum and $G$ the Newton's constant.
Due to the emission of Hawking radiation, the PBH mass and spin diminish with the rate given by the following system of coupled equations
\begin{subequations} \label{eq:MEq}
\begin{align}
 \frac{d\Mbh}{dt} &= - F(\Mbh, a_*)\, \frac{1}{G^2 \Mbh^2}\,,\\
  \frac{da_*}{dt} &= - a_*\left[G(\Mbh, a_*) - 2F(\Mbh, a_*)\right]\, \frac{1}{G^2 \Mbh^3}\,,
\end{align}
\end{subequations}
where $F(\Mbh, a_*)$ and $G(\Mbh, a_*)$, denoted as evaporation functions, contain the dependence on all the degrees of freedom that can be emitted, see Refs.~\cite{MacGibbon:1990zk, MacGibbon:1991tj,Cheek:2021odj} for further details.
To compute these evaporation functions, we use the code FRIedmann Solver for Black Hole Evaporation in the Early-universe {\tt FRISBHEE}~\cite{Cheek:2022dbx}, whose main library \texttt{BHProp.py} is included in {\ULYSSES} version 2 for convenience.
For further numerical convenience, we consider \emph{grams} as the main units for the PBH mass. The $a_*$ parameter is dimensionless.

To describe in detail the cosmological evolution, we solve the following set of Friedmann equations for the comoving radiation ($\varrho_{\rm R} \equiv a^4 \rho_{\rm R}$) and PBHs ($\varrho_{\rm BH} \equiv a^3 \rho_{\rm BH}$) energy densities to be solved together with the BH evolution equations, Eqs.~\eqref{eq:MEq}
\begin{subequations}\label{eq:UnEv},
\begin{align}
 	\frac{d\varrho_{\rm R}}{d \xi} &= -\frac{F_{\rm SM}(\Mbh, a_*)}{F(\Mbh, a_*)}\frac{1}{H}\frac{d\ln\Mbh}{dt}a\varrho_{\rm BH}\,,\label{eq:UnEvRad}\\ 
 \frac{d\varrho_{\rm BH}}{d \xi} &=\frac{1}{H}\frac{d\ln\Mbh}{dt} \varrho_{\rm BH}\,,\\
 H^2 &=\frac{8\pi G}{3} \left(\varrho_{\rm BH} 10^{-3x}+\varrho_{\rm R} 10^{-4x}\right)\,,
\end{align}
\end{subequations}
where $\xi \equiv \log_{10} (a/a_0)$ is the logarithm in base 10 of the scale factor $a$, $a_0$ the initial scale factor taken to equal 1, $H$ the Hubble rate.
Note that we evolve with respect to the dimensionless parameter $\xi$ instead of $z=M_1/T$ since entropy is not conserved throughout the evaporation, and thus we require a different independent variable.  

Assuming that the PBH formation occurs in a radiation-dominated era, we have that the initial PBH mass is related to the particle horizon mass as~\cite{Carr:2020gox}
\begin{align}\label{eq:Min}
 \Mbh^{\rm in} = \frac{4\pi}{3} \gamma\frac{\rho_i}{H_{\rm in}^3}\,,
\end{align}
with $\gamma=(1/\sqrt{3})^3$ the gravitational collapse factor, and $H_{\rm in}$ the Hubble parameter at the moment of PBH formation. 
Thus, by fixing the initial PBH mass, we define the initial conditions of the thermal plasma.
The initial PBH population is determined in  {\ULYSSES} via the dimensionless $\beta^\prime$ parameter, defined as
\begin{align}\label{eq:betap}
 \varrho^{\rm in}_{\rm BH}&=\frac{\beta^\prime}{\gamma^{1/2}-\beta^\prime}\varrho^{\rm in}_{\rm R} =\frac{\beta^\prime}{\gamma^{1/2}-\beta^\prime}\frac{\pi^2}{30} g_*(T_{\rm in})T_{\rm in}^4\,,
\end{align}
where $\varrho^{\rm in}_{\rm R}$ corresponds to the initial SM radiation energy density, determined by the initial PBH mass via Eq.~\eqref{eq:Min}, using that $H_{\rm in}\propto T_{\rm in}^2$ during the initial radiation dominated era.
Moreover, in this model, we assume a monochromatic PBH mass distribution, i.e., all black holes possess the same initial mass  $\Mbh^{\rm in}$ and spin $a_\star^{\rm in}$.
For later convenience, we also consider the explicit evolution of the SM thermal plasma temperature, $T$
\begin{equation} \label{eq:TUev}
  \frac{dT}{d \xi} = -\frac{T}{\Delta} \left\{1 -  \frac{\gs(T)}{\gss(T)} \frac{ 1}{4\varrho_{\rm R}}\frac{d\varrho_{\rm R}}{d \xi} \right\}\,,
\end{equation}
where $\Delta$ describes the change on the effective number of degrees of freedom $\gss(T)$ in Eq.~\eqref{eq:UnEvRad} 
\begin{equation}
\Delta \equiv 1 + \frac{T}{3 \gss(T)}\frac{d\gss(T)}{dT}\,.
\end{equation}
To determine the final baryon-to-photon ratio, we consider the momentum-integrated 
Boltzmann equations for the comoving thermal (${N}_{N}^{\rm TH}$) and non-thermal (${N}_{N}^{\rm BH}$) RHN densities~\cite{Perez-Gonzalez:2020vnz, Bernal:2022pue} 
\begin{subequations}\label{eq:BEright-handedn}
\begin{align}
	\frac{d{N}_{N}^{\rm TH}}{d \xi} &= -({N}_{N}^{\rm TH}-{ N}_{N}^{\rm eq})\frac{\Gamma_{N}^T}{H}\,,\label{eq:BERH-TH}\\
	\frac{d{ N}_{N}^{\rm BH}}{d\xi} &= -{ N}_{N}^{\rm BH}\frac{\Gamma_{N_1}^{\rm BH}}{H}+ { N}_{\rm BH} \frac{\Gamma_{{\rm BH}\to N_1}}{H}\,,\label{eq:BERH-BH}
\end{align}
\end{subequations}
where $\Gamma_{N_1}^T$, $\Gamma_{N_1}^{\rm BH}$ are the decay widths corrected by an inverse time dilation factor averaged over the plasma and BH temperature, respectively,
\begin{equation}\label{eq:GBH}
	\Gamma_{N_1}^{\rm T, BH} \equiv \left\langle\frac{M_1}{E_{N_1}}\right\rangle_{\rm T, BH} \Gamma_{N_1}\,,
\end{equation}
and $\Gamma_{N_1}$ is the RHN decay width. 
To address the generation of RHNs from the PBH density, we have included a source term in
\equaref{eq:BERH-BH} equal to the comoving PBH number density, ${ N}_{\rm BH}\equiv \varrho_{\rm BH}/\Mbh$, times $\Gamma_{{\rm BH}\to N_1}$, the total RHN emission rate per BH; see Ref.~\cite{Perez-Gonzalez:2020vnz}.
The equation for the $B-L$ asymmetry, ${ N}_{B-L}$, is
\begin{equation}\label{eq:BElep}
	\frac{d{ N}_{B-L}}{d \xi}= \frac{\epsilon}{H}\left[({ N}_{N}^{\rm TH}-{ N}_{N}^{\rm eq})\Gamma_{N_1}^T + { N}_{N}^{\rm BH} \Gamma_{N_1}^{\rm BH}\right] - \frac{1}{H}\left(\frac{1}{2}\Gamma_{N_1}^T { N}_{N}^{\rm eq}+\gamma\right)\frac{{ N}_{B-L}}{{ N}_{\ell}^{\rm eq}}\,,
\end{equation}
with ${ N}_{\ell}^{\rm eq}$ being the lepton equilibrium abundance. The term proportional to ${N}_{B-L}$ corresponds to the washout processes, including the $\Delta L=2$ interactions. 
After obtaining the $B-L$ number density, we similarly obtain $\eta_B$ as for other leptogenesis scenarios.
An important difference with the other models in ULYSSES should be noted here. 
The RHN neutrino abundances, both equilibrium and out-of-equilibrium, for thermal and PBH sources are normalized with respect to the \emph{initial} photon density, $n_{\gamma}^{\rm in} = 2\zeta(3)  T_{\rm in}^3/\pi^2$.

In  {\ULYSSES} version 2, we solve the system of equations Eqs.~\eqref{eq:MEq}, \eqref{eq:UnEv}, \eqref{eq:BEright-handedn}, \eqref{eq:BElep} together with the equation for the plasma temperature, Eq.~\eqref{eq:TUev}.
The code containing the mentioned system of equations and their solution is \texttt{etabPBH.py}, with the shortcut name being \texttt{1BE1F$\_$PBH}, and we provide an example parameter card; details of their usage are found in \secref{sec:usage}.
Since the time-evolution equations for the PBH mass and spin become quite stiff when the evaporation enters the final stages, we have implemented an iterative approach to evolve until the PBH mass reaches the Planck Mass, the point at which we stop the evolution.
We solve the equations from the initial black hole mass until 1\% of the initial value and then take the found solutions as initial conditions for a new iteration.
This is done until the PBH mass arrives at the Planck scale.
If thermal leptogenesis occurs after the PBH evaporation, we have added a second set of BEs, which is solved using the solutions obtained after properly evolving the PBH particle production. Finally, the user can modify the initial abundances of particles by adjusting the \texttt{y0} array in the model file.
\section{Installation}\label{sec:install}
\noindent The code is hosted on \url{https://github.com/earlyuniverse/ulysses}. 
Once the git repository is pulled,
the basic installation steps are shown in \listing{lst:install}. In
addition, releases are packaged and available to install with {\sc pip} from
\url{https://pypi.org/}.

\begin{minipage}{\linewidth}
\lstset{label={lst:install}}
\lstset{caption={Minimal installation steps.}}
\begin{bash}
# Installation from within the source tree
git clone https://github.com/earlyuniverse/ulysses.git
cd ulysses
pip install . --user

# Installation with pip or pip3 from pypi.org
pip install ulysses --user
\end{bash}
\end{minipage}
\subsection{Core dependencies}
\noindent The code is written in Python3 and heavily uses the widely available modules {\sc
NumPy}~\cite{oliphant2006guide,van2011numpy} and {\sc SciPy}~\cite{virtanen2020scipy} packages \footnote{We note that outdated versions of {\sc
NumPy} and {\sc SciPy} may lead to numerical instabilities, especially for {\ttfamily 1BE1F$\_$Case2}, {\ttfamily 1BE1F$\_$Case3} and {\ttfamily 1BE1F$\_$Case4}, and thus we recommend that the user upgrade to the latest, up-to-date versions.
}.
We accelerate the computation with the just-in-time compiler provided by {\sc
Numba}~\cite{lam2015numba} where meaningful.  At its core,
{\ULYSSES} solves a set of coupled differential equations. To undertake this
task we use \texttt{solve}$\_$\texttt{ivp} and \texttt{odeintw}  \cite{odeintw}. The former is a standard Python package for solving initial value problems for ordinary differential equations, while the latter provides a wrapper of
{\ttfamily scipy.integrate.odeint} that allows it to handle complex and matrix differential
equations; it is redistributed with {\ULYSSES} and does not need to be downloaded separately. These dependencies for {\ULYSSES} are automatically
resolved during the install process with {\sc pip}. They provide the minimal
functionality for solving Boltzmann equations for a given point in the model
parameter space. In {\ULYSSES} version 2, Python packages {\sc tqdm} and {\sc termcolor} must also be pip-installed. 

\begin{table}[t!]
    \centering
    \begin{tabular}{llcc}
        \toprule
        Parameter                                                         & Variable name      & Default               & Unit            \\
        \midrule
        Higgs vacuum expectation value,                                     $v$                & {\texttt{vev}}      & $174.0$               & [GeV]           \\
        Higgs mass,                                    $M_H$              & {\texttt{mhiggs}}   & $125.35$               & [GeV]           \\
        Z boson mass,                                  $M_{Z}$            & {\texttt{mz}}       & $91.1876$             & [GeV]           \\
        Planck mass,                                   $M_{\text{PL}}$    & {\texttt{mplanck}}  & $1.22\times10^{19}$   & [GeV]           \\
        Neutrino cosmological mass,                    $m_*$              & {\texttt{mstar}}    & $10^{-12}$            & [GeV]           \\
        Degrees of freedom,                            $g_*$               & {\texttt{gstar}}    & $106.75$              &                 \\
        Solar mass squared splitting,                   $\Delta m^2_{\odot}$ & {\texttt{m2solar}}  & $7.42\times10^{-23}$   & [GeV$\strut^2$] \\
        Atm. mass squared splitting (normal),          $\Delta m^2_{\text{atm}}$ & {\texttt{m2atm}}    & $2.515\times10^{-21}$ & [GeV$\strut^2$] \\
        Atm. mass squared splitting (inverted),        $\Delta m^2_{\text{atm, inv}}$ & {\texttt{m2atminv}} & $2.498\times10^{-21}$ & [GeV$\strut^2$] \\
        \bottomrule 
    \end{tabular}
    \caption{Overview of global parameters and their default values. Neutrino mass squared splittings are taken from the central values of NuFIT 5.1 global fit data (without atmospheric data from Super-Kamiokande) \cite{Esteban:2020cvm}.}
    \label{tab:const}
\end{table}

\section{Usage of {\ULYSSES} version 2}\label{sec:usage}

\subsection{The model parameters}\label{sec:params}
\noindent All global constants are defined in the {\ttfamily \_\_init\_\_} function of the base
class and they are shown in \tabref{tab:const}. We allow the user to set their values via the standard Python keyword argument
formalism using the variable names shown in the second column
of~\tabref{tab:const}.
The required input from the user is the set of model parameters which stems from the
Casas-Ibarra parametrisation of the Yukawa matrix $Y$, as shown below \cite{Casas:2001sr}:
\begin{equation}\label{eq:CandI}
Y=\frac{1}{v}U\sqrt{\hat{{m}}_{\nu}}R^T\sqrt{M_R}\,.
\end{equation}
{where
$v = 174\,\text{GeV}$ is the Higgs's vacuum expectation value, $U$ is the $3\times3$ unitary Pontecorvo-Maki-Nakagawa-Sakata (PMNS) regulating the neutrino (lepton) mixing, $\hat{m}_\nu = \text{diag}(m_1,m_2,m_3)$ is the diagonal light neutrino mass matrix, $R$
is a  $3\times 3$ complex orthogonal matrix and $M_R = {\rm diag}(M_1,M_2,M_3)$ is the diagonal mass matrix of the RHNs.}
We apply the Particle Data Group convention \cite{Tanabashi:2018oca} to parameterise the PMNS matrix:
\begin{equation}
\begin{aligned}
U =&\begin{pmatrix}
1 & 0 & 0 \\
0 & c_{23} & s_{23}  \\
0 &- s_{23} & c_{23} 
\end{pmatrix}
\begin{pmatrix}
c_{13} & 0 & s_{13}e^{-i\delta} \\
0 & 1 & 0 \\
-s_{13}e^{i\delta} &0 & c_{13}
\end{pmatrix}
\begin{pmatrix}
c_{12} & s_{12} & 0\\
-s_{12} & c_{12} & 0\\
0 & 0 & 1
\end{pmatrix}
\begin{pmatrix}
1 & 0 & 0\\
0&e^{i\frac{\alpha_{21}}{2}} & 0\\
0 & 0 &  e^{i\frac{\alpha_{31}}{2}}
\end{pmatrix}\,,
\end{aligned}
\end{equation}
 where $c_{ij} \equiv \cos\theta_{ij}$, $s_{ij} \equiv \sin\theta_{ij}$, $\delta$ is the Dirac  phase and $\alpha_{21}$, $\alpha_{31}$ are the Majorana phases \cite{Bilenky:1980cx} {which, in general, can vary between $0 \leq\alpha_{21}, \alpha_{31} \leq 2\pi$.}
The $R$-matrix can be written in the following form:
\begin{equation}\label{eq:Rmatrix}
R=\begin{pmatrix}
1 & 0 & 0 \\
0 & c_{\omega_{1}} & s_{\omega_{1}} \\
0 &- s_{\omega_{1}} & c_{\omega_{1}} 
\end{pmatrix}
\begin{pmatrix}
c_{\omega_{2}} & 0 & s_{\omega_{2}} \\
0 & 1 & 0\\
-s_{\omega_{2}} & 0 & c_{\omega_{2}} 
\end{pmatrix}\\
\begin{pmatrix}
c_{\omega_{3}} & s_{\omega_{3}} & 0\\
-s_{\omega_{3}} & c_{\omega_{3}} & 0\\
0 & 0 & 1
\end{pmatrix}\,,
\end{equation}

 \begin{table}
     \centering
     \begin{tabular}{lc|cr}
     \toprule
         Parameter & Unit & \multicolumn{2}{c}{Code input example} \\ \hline
         $\delta~$                       & $\left[^\circ\right]$      &{\ttm delta} &{\ttm 213.70} \\
         $\alpha_{21}~$                  & $\left[^\circ\right]$      &{\ttm a21} & {\ttm 81.60}\\
         $\alpha_{31}~$                  & $\left[^\circ\right]$      &{\ttm a31}  & {\ttm 476.70}\\
         $\theta_{23}~$                  & $\left[^\circ\right]$      &{\ttm t23}  & {\ttm 48.58}\\
         $\theta_{12}~$                  & $\left[^\circ\right]$      &{\ttm t12}   & {\ttm 33.63}\\
         $\theta_{13}~$                  & $\left[^\circ\right]$      &{\ttm t13}   & {\ttm 8.52}\\ \hline
         $ x_{1}~$                       & $\left[^\circ\right]$      &{\ttm x1}    & {\ttm 90.00}\\
         $ y_{1}~$                       & $\left[^\circ\right]$      &{\ttm y1}    & {\ttm -120.00}\\
         $ x_2~$                           & $\left[^\circ\right]$      &{\ttm x2} & {\ttm 87.00}\\
         $ y_{2}~$                       & $\left[^\circ\right]$      &{\ttm y2}    & {\ttm 0.00}\\
         $ x_3~$                           & $\left[^\circ\right]$      &{\ttm x3}    & {\ttm 180.00}\\
         $ y_{3}~$                       & $\left[^\circ\right]$      &{\ttm y3}   & {\ttm -120.00}\\ \hline
         $\log_{10}\left(m_{1/3}\right)$ & $\left[\mathrm{eV}\right]$ &{\ttm m}     & {\ttm -1.10}\\
         $\log_{10}\left(M_1\right)$    & [$\GeV$]                    &{\ttm M1}    & {\ttm 12.10}\\
         $\log_{10}\left(M_2\right)$    & [$\GeV$]                    &{\ttm M2}    & {\ttm 12.60}\\
         $\log_{10}\left(M_3\right)$    & [$\GeV$]                    &{\ttm M3}    & {\ttm 13.00}\\
             \bottomrule
     \end{tabular}
     \caption{Overview of the input parameters in the Casas-Ibarra parametrisation.}
     \label{tab:params}
\end{table}
where $c_{\omega_{i}} \equiv \cos\omega_{i}$, $s_{\omega_{i}} \equiv \sin\omega_{i}$ 
 and the complex angles are given by $\omega_{i} \equiv x_{i}+iy_{i}$ for $x, y$  free, real parameters. 
The $R$-matrix in Eq.~\eqref{eq:Rmatrix} have det$(R) = 1$.
Often, in the literature, a phase factor
$\varphi = \pm 1$ is included in the definition of certain elements
of the matrix $R$ to allow for the both cases det$(R) = \pm 1$.
However, one can extend the range of values of the Majorana phases to 
$0 \leq\alpha_{21}, \alpha_{31} \leq 4\pi$ to effectively account for both cases of $\text{det}(R) = \pm\, 1$ and consider, in this way, the same full set of $R$ and Yukawa matrices \cite{Molinaro:2008rg}.

The Casas-Ibarra parameters (and their units) may be input by the user in the code and an  example is given in \tabref{tab:params}. Specifically, assigning a value to the code variables named {\ttm delta}, {\ttm a21}, {\ttm a31}, {\ttm t23}, {\ttm t12}, {\ttm t13}, {\ttm x1}, {\ttm y1}, {\ttm x2}, {\ttm y2}, {\ttm x3}, {\ttm y3}, {\ttm M1}, {\ttm M2}, {\ttm M3} and {\ttm m}, the user fixes, respectively, the PMNS phases and angles in degrees, the real and imaginary parts of the three complex angles of the $R$-matrix in degrees, the three RHN masses in GeV and of the lightest neutrino mass (that is either $m_1$ or $m_3$, depending on the ordering of the light neutrino masses) in eV. In contrast, the two heavier neutrino masses are fixed at the best-fit values using the global fit data on solar and atmospheric mass squared differences \cite{Esteban:2020cvm}, the values of which, if necessary, can be changed directly by the user in \texttt{ulsbase.py}.
Since the last release, we have updated the neutrino parameters to 
the NuFIT 5.1 global fit central values \cite{Esteban:2020cvm}. We stress that, as an input, the user fixes the logarithm in base 10 of the masses of the RHNs and the lightest active neutrino. In particular, using the example in \tabref{tab:params}, which is also given separately in \listing{lst:examplecalc1} in the form of a parameter card, the lightest active neutrino mass is fixed at $m_{1}=10^{-1.1}$ eV and the RHN masses at $M_{1,\,2,\,3} = 10^{12.1,\,12.6,\,13}$ GeV.

The Casas-Ibarra parametrisation is one popular parametrisation of the Yukawa matrix that guarantees the correct prediction of the observed pattern of light neutrino masses and mixing. However,
{\ULYSSES} also allows the user to provide their own Yukawa matrix in polar
coordinates and calculate the resultant baryon asymmetry. We note that, in the latter option, the user will need to ensure independently that the oscillation data are satisfied.
The input logic is such that each element of the Yukawa matrix, $Y_{ij}$, is determined by two independent parameters {\ttm Yij\_mag} and {\ttm Yij\_phs},
which are the absolute magnitude and phase (the polar coordinates) of the Yukawa entry, respectively:
\begin{equation}
\begin{aligned}
    Y_{ij}& =  {\mathtt{ Yij\_mag}}\cdot\exp\left(i~\mathtt{Yij\_phs}\right)\,.
\end{aligned}
\end{equation}
An example input card {for a generic Yukawa parametrisation in polar coordinates} is shown in~\listing{lst:examplecalcYukawa}.

The code for PBH-induced leptogenesis requires the user to specify also the parameters $a_*$, $\beta^\prime$ and $ \Mbh^{\rm in} $ introduced in \secref{sec:PBH}. Therefore, the user, in addition to the parameters of the type-I seesaw model described above, needs to fix also the parameters $a_*$, $\beta^\prime$ and $ \Mbh^{\rm in} $ by assigning a value to each code variables named {\ttm aPBHi}, {\ttm bPBHi} and  {\ttm MPBHi}  (see further in \secref{sec:examples} for a specific parameter card and related example). {\ttm MPBHi} is the initial mass of the black holes in grams (monochromatic mass distribution) in logarithm of  base 10, {\ttm aPBHi} is the dimensionless parameter related to the spin and 
takes values between $0$ (spinless) and $1$ (maximally spinning) and, finally, {\ttm bPBHi} is related to the initial number density of black holes (in logarithm of  base 10).

 \begin{minipage}{.4\textwidth} %
 \lstset{label={lst:examplecalc1}}
 \lstset{caption={Example of an input parameter card using the Casas-Ibarra parametrisation.}}
 \begin{python}
 m       -1.10
 M1      12.10
 M2      12.60
 M3      13.00
 delta  213.70
 a21     81.60
 a31    476.70
 x1      90.00
 x2      87.00
 x3     180.00
 y1    -120.00
 y2       0.00
 y3    -120.00
 t12     33.63
 t13      8.52
 t23     49.58
 \end{python}
 \end{minipage} %
 \hspace{2cm}
\begin{minipage}{.4\textwidth} %

\lstset{label={lst:examplecalcYukawa}}
\lstset{caption={Example of an input parameter card for generic Yukawas and RHN masses.}}
\begin{python}
Y11_mag  0.01
Y12_mag  0.01 
Y13_mag  0.01
Y21_mag  0.01
Y22_mag  0.03
Y23_mag  0.05
Y31_mag  0.01
Y32_mag  0.03
Y33_mag  0.05
Y11_phs -1.11
Y12_phs  2.89
Y13_phs  1.32
Y21_phs  2.88
Y22_phs -0.23
Y23_phs -1.80
Y31_phs -1.72
Y32_phs  2.96
Y33_phs  1.39
M1       12.0
M2       12.5
M3       13.0
\end{python}
\end{minipage}

\subsection{Functionalities}\label{sec:func}
\noindent For convenience, we ship four runtime scripts which use the {\ULYSSES} module for
the evaluation of $\eta_B$ at a single point, as well as in one-dimensional and in
multi-dimensional parameter space explorations:
\begin{itemize}
    \item {\ttm uls-calc}
    \item {\ttm uls-scan}
        \item {\ttm uls-nest}
        \item {\ttm uls-scan2D}
\end{itemize}
The first three of the functionalities listed above were already shipped with {\ULYSSES} version 1 \cite{Granelli:2020pim}, while the new functionality shipped with version 2 is {\ttm uls-scan2D},
which generalises {\ttm uls-scan} to two dimensions. We provide some examples of the usage of the functionalities within {\ULYSSES} version 2 in the next subsections.

\subsection{Examples}\label{sec:examples}
\noindent To display the pre-provided BEs, including those discussed in \secref{sec:plugins}, and the strings needed to load them from the command line 
the user can call:
\begin{bash}
# display list of available models
uls-models
\end{bash}

In addition to those BE codes shipped with {\ULYSSES} version 1, we have included new codes with their own 
 ``shortcut names''. For the cases of thermal leptogenesis discussed in \secref{sec:TL}, the  ``shortcut names'' for 
 Cases 2 to 4 are  \texttt{1BE1F$\_$Case2}, \texttt{1BE1F$\_$Case3}, \texttt{1BE1F$\_$Case4} respectively. An example call to \texttt{uls-calc} on \texttt{1BE1F$\_$Case4}  is shown below:
 \begin{bash}
# call 1BE1F_Case4 on input card 1N1F.dat and output 1BE1F_Case4.pdf
uls-calc -m 1BE1F_Case4 examples/1N1F.dat -o 1BE1F_Case4.pdf
\end{bash}
which returns the baryon asymmetry (in terms of the baryon-to-photon ratio, the baryonic yield and the 
baryonic density) and a plot  (1BE1F$\_$Case4.pdf) showing the time evolution of the lepton asymmetry and baryon-to-photon ratio.

For ARS leptogenesis, there are two modules with shortcut names \texttt{BEARS} and \texttt{BEARS$\_$INTERP}, which use temperature-independent and
 temperature-dependent rates, respectively. The latter module also takes into account non-relativistic corrections to the LNV rates as detailed in Ref.~\cite{Hernandez:2022ivz}. An example call for this code is
  \begin{bash}
# call ARS BE with temperature independent rates on input card pars.dat
uls-calc -m BEARS examples/2RHNosc.dat 
\end{bash}
where the output displayed on the terminal consists of the Yukawa matrix and baryon asymmetry given
in terms of $\eta_B$, $Y_B$ and $\Omega_B$. Note  that, to use the temperature-dependent rates in the above command, \texttt{BEARS$\_$INTERP} should be used instead of \texttt{BEARS}.

In the above example, the  input
card is shown in \listing{lst:exampleARS}. In this case, the mass splittings are very small, and the oscillation length, as discussed in \secref{sec:ARS}, is
$z_\text{osc}>0.1$ and no stitching of the solutions is required. There is a second example card, 
 named \texttt{2RHNosc$\_$examplestitch.dat}, which has a larger mass squared splitting between the two right-handed neutrinos resulting in $z_\text{osc}<0.1$. 
The integration time can be longer in this case, so the code allows the user to specify where the stitch should occur:
  \begin{bash}
# call BE_ARS on input card pars.dat
uls-calc -m BEARS examples/2RHNosc_examplestitch.dat --zcut 0.6
\end{bash}
where, in the above case, the cut is chosen to be at $z_\mathrm{cut} = 0.6$. The example input card used in the above example is shown in \listing{lst:exampleARSstitch}. The ARS code also automatically outputs a plot of the absolute magnitudes of the chemical potentials as a function of $z$. 

\begin{figure}[t!]
    \centering
    \includegraphics[width=\textwidth]{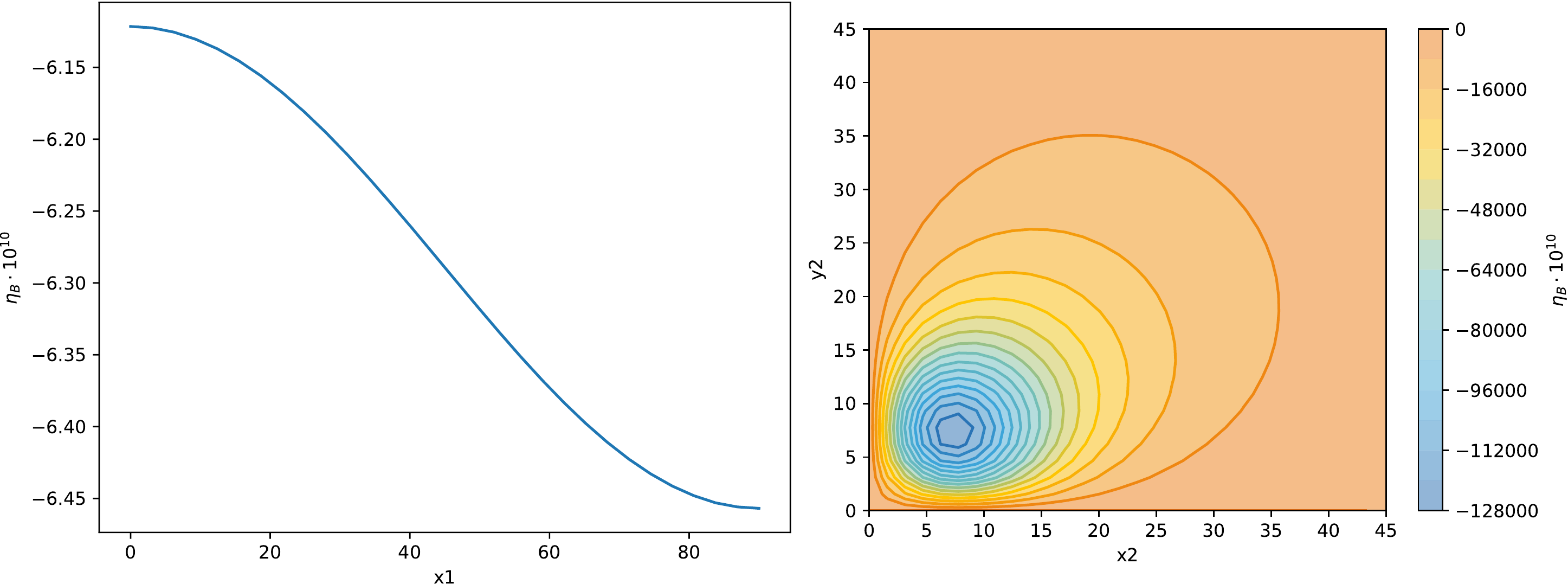}
    \caption{On the left is the one-dimensional scan output of \texttt{1BE1F}$\_$\texttt{PBH.py} on the parameter card \texttt{PBH.dat} in \listing{lst:examplePBH}, while on the right is the contour plot output from \texttt{uls-scan2D} using the input card \texttt{1N1F}$\_$\texttt{2Dscan.dat} in \listing{lst:examplescan2D}.}
    \label{fig:contour}
\end{figure}

To call PBH-induced leptogenesis requires an input card not only with the usual Casas-Ibarra parametrisation but also 
with the PBH parameters ($a_*$, $\beta^\prime$ and $ \Mbh^{\rm in} $), as detailed in \secref{sec:PBH} and \secref{sec:params}. An example
input card shipped with {\ULYSSES} version 2 is shown in \listing{lst:examplePBH}. 
A code example where a one-dimensional scan in variable $x_1$ is performed using the parameters of \listing{lst:examplePBH} is given below:
 \begin{bash}
# call 1BE1F_PBH on input card PBH.dat
uls-scan -m 1BE1F_PBH examples/PBH.dat -o PBHscan.pdf
\end{bash}
The one-dimensional scan's illustrative output is given in the left panel of \figref{fig:contour}.

\begin{minipage}{.4\textwidth} %
\lstset{label={lst:exampleARS}}
\lstset{caption={Example of an input card for ARS leptogenesis ({\ttfamily 2RHNosc.dat}).}}
\begin{python}
m -100.
M1 -100.
M2 0.0
M3 4.34294e-9
x1 45.
y1 40.107
x2 0.
y2 0.
x3 0.
y3 0.
delta 221.
a21 322.
a31 0.
t12 33.8486
t13 8.60954
t23 48.5904

\end{python}
\end{minipage} %
\hspace{2cm}
\begin{minipage}{.4\textwidth} %

\lstset{label={lst:exampleARSstitch}}
\lstset{caption={Example of an input card for ARS leptogenesis where the stitch is required ({\ttfamily 2RHNosc\_examplestitch.dat}).}}
\begin{python}
m -100.
M1 -100.
M2 0.0
M3 9.34294e-9
x1 45.
y1 40.107
x2 0.
y2 0.
x3 0.
y3 0.
delta 221.
a21 322.
a31 0.
t12 33.8486
t13 8.60954
t23 48.5904
\end{python}
\end{minipage}\\
 
\noindent An example of applying {\ttm uls-scan2D} is given below:
\begin{bash}
# Use one of the built-in plugins
uls-scan2D -m 1BE1F examples/1N1F_2Dscan.dat -o example_2D.pdf
\end{bash}
where the input parameter card \texttt{1N1F}$\_${\texttt{2Dscan.dat}}, shown in \listing{lst:examplescan2D}, let the parameters 
$x_2$ and $y_2$ vary within the range $[0^\circ\,, 45^\circ]$. The code saves a pdf file depicting a contour plot of $\eta_B\times 10^{10}$ as a function of varied parameters, as shown in the right panel of \figref{fig:contour}. If the user wishes to obtain a text file with the numerical output, the following command can be used:
\begin{bash}
# Use one of the built-in plugins
uls-scan2D -m 1BE1F examples/1N1F_2Dscan.dat -o example_2D.txt
\end{bash}
saving in the first and second columns of the output text file the values of the first and second varied parameters, respectively, while in the third column the calculated $\eta_{B}\times 10^{10}$.

\begin{minipage}{0.4\textwidth} %
\lstset{label={lst:examplePBH}}
\lstset{caption={Example of an input card (\tt PBH.dat}) for the PBH-assisted leptogenesis model.}
\begin{python}
m -0.576754
M1 14.0000
M2 14.4771
M3 14.7785
x1 0.0 90.0
y1 0.
x2 45.0000
y2 25.2495
x3 0.
y3 0.
delta 194.
a21 0.
a31 0.
t12 33.44
t13 8.57
t23 49.2
MPBHi 0.0
aPBHi 0.0 
bPBHi -5.68
\end{python}
\end{minipage} 
\hspace{2cm}
\begin{minipage}{0.4\textwidth} %

\lstset{label={lst:examplescan2D}}
\lstset{caption={Example of an input parameter card ({\tt 1N1F\_2Dscan.dat}) for {\ttfamily uls-scan2D} using the Casas-Ibarra parametrisation.}}
\begin{python}
m -100
M1 14
M2 15
M3 16
x1 180
y1 1.4
x2 0. 45.
y2 0. 45.
x3 180
y3 11
delta 217.
a21 0.
a31 0
t23 49.7
t12 33.82
t13 8.610
\end{python}
\end{minipage} %
\\

\section{Summary and Discussion}\label{sec:conclusions}
\noindent In this second release of {\ULYSSES} we have implemented the Boltzmann equations for the complete phase-space evolution of
thermal leptogenesis, based on the work of Ref.~\cite{Hahn-Woernle:2009jyb}, the equations for non-resonant leptogenesis in the context of a primordial black hole dominated early Universe (see Refs.~\cite{Perez-Gonzalez:2020vnz,Bernal:2022pue})
and, finally, the kinetic equations for leptogenesis via oscillations \cite{Akhmedov:1998qx,Asaka:2005pn} based on a notation similar to that of Ref.~\cite{Hernandez:2022ivz} (see also Refs.~\cite{Hernandez:2016kel,Abada:2018oly}). The functionality of {\ULYSSES} 
has been expanded with the facility for a two-dimensional scan. As stated in the first version of the manual \cite{Granelli:2020pim}, we view this as a community project and invite users to add and share their plugins with others. This can be done via issues and pull requests on our GitHub repository.

\section*{Acknowledgements}
\noindent We would like to thank Roberta Calabrese and Serguey T.~Petcov for useful feedback on the output of the code for PBH and ARS leptogenesis, respectively, and we would like to thank Stefan Sandner and Dave Tucker-Smith for helpful discussions.
This work used the DiRAC@Durham facility managed by the Institute for Computational Cosmology on behalf of the STFC DiRAC HPC Facility (\href{www.dirac.ac.uk}{www.dirac.ac.uk}). The equipment was funded by BEIS capital funding via STFC capital grants ST/P002293/1, ST/R002371/1 and ST/S002502/1; Durham University and STFC operations grant ST/R000832/1. DiRAC is part of the National e-Infrastructure. 
This work has made use of the Hamilton HPC Service of Durham University. The work of BS is supported by Research Corporation for Science Advancement through Cottrell Scholar Grant \#27632.

\bibliographystyle{JHEP}
\bibliography{leptobib}

\end{document}